\documentclass[aps, prb, twocolumn, reprint, superscriptaddress, floatfix, citeautoscript, longbibliography]{revtex4-1}
\pdfoutput=1
\usepackage{graphicx}
\usepackage{siunitx}
\usepackage{dcolumn} 
\usepackage{bm} 
\usepackage{chemformula} 
\usepackage{comment}
\usepackage{braket}

\usepackage{color}

\usepackage[
   unicode,
   colorlinks = true,
   allcolors = blue,
]{hyperref}

\usepackage{cleveref}

\begin{document} 
\newcommand{\bNMR}{$\beta$-NMR}
\newcommand{\bnmr}{$\beta$-NMR}
\newcommand{\bNQR}{$\beta$-NQR}
\newcommand{\ie}{{\it i.e.,}}
\newcommand{\eg}{{\it e.g.,}}
\newcommand{\etc}{{\it etc.}}
\newcommand{\etal}{{\it et al.}}
\newcommand{\ibid}{{\it ibid.}}
\newcommand{\dC}{$^{\circ}$C}
\newcommand{\Fig}[1]{{\mbox{Fig.~#1}}}
\newcommand{\Table}[1]{{\mbox{Table~#1}}}
\newcommand{\eli}{$^{8}$Li}
\newcommand{\elip}{$^{8}$Li$^{+}$}
\newcommand{\elin}{$^{8}$Li$^{0}$}
\newcommand{\lip}{Li$^{+}$}
\newcommand{\intp}{Li$_{\mathrm{i}}^{\mathrm{O}+}$}
\newcommand{\inth}{Li$_{\mathrm{i}}^{\mathrm{O}}$}
\newcommand{\intLi}{Li$_{\mathrm{i}}$}
\newcommand{\Tp}{Li$_{\mathrm{T}}^{+}$}
\newcommand{\Tint}{Li$_{\mathrm{T}}$}
\newcommand{\subn}{Li$_{\mathrm{Zn}}^{-}$}
\newcommand{\sub}{Li$_{\mathrm{Zn}}^{0}$}
\newcommand{\subLi}{Li$_{\mathrm{Zn}}$}
\newcommand{\ks}{\mbox{\it K}}
\newcommand{\tone}{\mbox{\it T}$_{1}$}
\newcommand{\kr}{\mbox{$\cal K$}}
\newcommand{\nuq}{\mbox{$\nu_{\it q}$}}
\newcommand{\uSR}{$\mu$SR}
\newcommand{\usr}{$\mu$SR}
\newcommand{\mup}{$\mu^+$}
\newcommand{\lao}{LaAlO$_3$}
\newcommand{\T}{$\mathrm{T}$}
\newcommand{\rom}[1]{\uppercase\expandafter{\romannumeral #1\relax}}
\newcommand{\np}{$n^+$}
\newcommand{\VO}{V$_{\mathrm{O}}$}
\newcommand{\VZn}{V$_{\mathrm{Zn}}$}
\newcommand{\VZnn}{V$_{\mathrm{Zn}}^{-2}$}
\newcommand{\ZnI}{Zn$_{\mathrm{I}}$}
\newcommand{\ZnIp}{Zn$_{\mathrm{I}}^{+2}$}
\title{Nuclear magnetic resonance of ion implanted $^8$Li in ZnO }

\author{Jonah R. Adelman}
\email[Email: ]{jradelman@berkeley.edu}
\altaffiliation{Current address: Department of Chemistry, University of California, Berkeley, CA 94720, USA}
\affiliation{Department of Chemistry, University of British Columbia, Vancouver, BC V6T~1Z1, Canada}

\author{Derek Fujimoto}
\affiliation{Stewart Blusson Quantum Matter Institute, University of British Columbia, Vancouver, BC V6T~1Z4, Canada}
\affiliation{Department of Physics and Astronomy, University of British Columbia, Vancouver, BC V6T~1Z1, Canada}

\author{Martin H. Dehn}
\affiliation{Stewart Blusson Quantum Matter Institute, University of British Columbia, Vancouver, BC V6T~1Z4, Canada}
\affiliation{Department of Physics and Astronomy, University of British Columbia, Vancouver, BC V6T~1Z1, Canada}

\author{Sarah R. Dunsiger}
\affiliation{TRIUMF, 4004 Wesbrook Mall, Vancouver, BC V6T~2A3, Canada}
\affiliation{Department of Physics, Simon Fraser University, Burnaby, BC V5A~1S6, Canada}

\author{Victoria L. Karner}
\affiliation{Department of Chemistry, University of British Columbia, Vancouver, BC V6T~1Z1, Canada}
\affiliation{Stewart Blusson Quantum Matter Institute, University of British Columbia, Vancouver, BC V6T~1Z4, Canada}

\author{C. D. Philip Levy}
\affiliation{TRIUMF, 4004 Wesbrook Mall, Vancouver, BC V6T~2A3, Canada}

\author{Ruohong Li}
\affiliation{TRIUMF, 4004 Wesbrook Mall, Vancouver, BC V6T~2A3, Canada}

\author{Iain McKenzie}
\affiliation{TRIUMF, 4004 Wesbrook Mall, Vancouver, BC V6T~2A3, Canada}

\author{Ryan M. L. McFadden}
\altaffiliation{Current address: TRIUMF, 4004 Wesbrook Mall, Vancouver, BC V6T~2A3, Canada}
\affiliation{Department of Chemistry, University of British Columbia, Vancouver, BC V6T~1Z1, Canada}
\affiliation{Stewart Blusson Quantum Matter Institute, University of British Columbia, Vancouver, BC V6T~1Z4, Canada}

\author{Gerald D. Morris}
\affiliation{TRIUMF, 4004 Wesbrook Mall, Vancouver, BC V6T~2A3, Canada}

\author{Matthew R. Pearson}
\affiliation{TRIUMF, 4004 Wesbrook Mall, Vancouver, BC V6T~2A3, Canada}

\author{Monika Stachura}
\affiliation{TRIUMF, 4004 Wesbrook Mall, Vancouver, BC V6T~2A3, Canada}

\author{Edward Thoeng}
\affiliation{Department of Physics and Astronomy, University of British Columbia, Vancouver, BC V6T~1Z1, Canada}
\affiliation{TRIUMF, 4004 Wesbrook Mall, Vancouver, BC V6T~2A3, Canada}

\author{John O. Ticknor}
\affiliation{Department of Chemistry, University of British Columbia, Vancouver, BC V6T~1Z1, Canada}
\affiliation{Stewart Blusson Quantum Matter Institute, University of British Columbia, Vancouver, BC V6T~1Z4, Canada}

\author{Naoki Ohashi} 
\affiliation{National Institute for Materials Science (NIMS), 1-1 Namiki, Tsukuba 305-0044, Japan
}

\author{Kenji M. Kojima}
\affiliation{TRIUMF, 4004 Wesbrook Mall, Vancouver, BC V6T~2A3, Canada}

\author{W. Andrew MacFarlane}
\email[Email: ]{wam@chem.ubc.ca}
\affiliation{Department of Chemistry, University of British Columbia, Vancouver, BC V6T~1Z1, Canada}
\affiliation{Stewart Blusson Quantum Matter Institute, University of British Columbia, Vancouver, BC V6T~1Z4, Canada}
\affiliation{TRIUMF, 4004 Wesbrook Mall, Vancouver, BC V6T~2A3, Canada}

\date{\today} 

\begin{abstract} 
We report on the stability and magnetic state of ion implanted \eli{} in single crystals 
of the semiconductor ZnO using $\beta$-detected nuclear magnetic resonance. At ultradilute concentrations, 
the spectra reveal distinct Li sites from 7.6 to 400 K. Ionized shallow donor 
interstitial Li is stable across the entire temperature range, confirming its ability to 
self-compensate the acceptor character of its (Zn) substitutional counterpart. Above 300 K, 
spin-lattice relaxation indicates the onset of correlated local motion of interacting defects, and 
the spectra show a site change transition from disordered configurations to substitutional. Like the interstitial, 
the substitutional shows no resolved hyperfine splitting, 
indicating it is also fully ionized above 210 K. The electric field gradient at the interstitial \eli{} 
exhibits substantial temperature dependence with a power law typical of non-cubic metals.
\end{abstract}

\maketitle 

\section{Introduction} 

For many years the electronic properties of ZnO have held substantial practical appeal. 
It is a transparent wide bandgap (3.4 eV) semiconductor with a large exciton 
binding energy (60 meV) enabling potential room 
temperature optoelectronic applications\cite{JanottiRPP2009,LookMSEB2001}.
It can be doped magnetically, either by replacing the nonmagnetic $(3d^{10})$ Zn$^{2+}$ 
with a magnetic transition metal (or, more subtly, by intrinsic magnetic defects\cite{Esquinazi2020}) to 
produce a room temperature dilute magnetic semiconductor\cite{SharmaNM2003} useful in spintronics\cite{Botsch2019}.
It also exhibits many interesting surface effects including photocatalysis, 
adsorbate sensitive photoconductivity\cite{Gurwitz2014}, and a photogenerated metallic state\cite{Gierster2021}.  
Key to realizing the potential of this remarkable range of properties is understanding and 
control of the point defects (both intrinsic and extrinsic) of its ideal
hexagonal wurtzite structure\cite{McCluskey2009}. Particularly, for $p$-type doping this 
has proven difficult\cite{OzgurJAP2005}, but there are clear indications of progress in 
the production of LEDs\cite{Rahman2019} and even extremely high mobility 
epitaxial structures exhibiting the quantum Hall effect\cite{Falson2018}.

Consistent and stable hole doping of ZnO is an important challenge that 
has remained the main barrier to implementation in optoelectronic devices. 
Substitution of divalent Zn for a monovalent alkali is an approach that has been 
studied in some detail\cite{JanottiRPP2009,McCluskey2009}. While substitutional Li 
is an acceptor that compensates the natural $n$-type doping to some extent, this 
has not led to a reliable route to $p$-type ZnO. One reason for this is that Li is an 
amphoteric dopant: when interstitial it is a donor, so it can self-compensate. 
In addition to carrier doping, the magnetic aspect of Li defects is of interest 
for spin filtering\cite{Esquinazi2020}. In this context, detailed experimental characterization of the 
structure, stability, and magnetic state of isolated Li defects 
would be very useful, particularly as a test of theoretical calculations.

To this end, we implanted a low energy (20-25 keV) beam of the shortlived radioisotope \elip\ 
into high purity single crystals of ZnO and studied the resulting isolated 
implant via its nuclear magnetic resonance as reported by its radioactive 
beta decay (\bnmr)\cite{MacFarlaneSSNMR2015}. Closely related to muon spin 
rotation (\usr{}) which has yielded an intricate understanding of isolated 
hydrogen impurities in ZnO \cite{ShimomuraPRL2002,CoxPRL2001}, \bnmr\ enables complementary 
investigations of other light isotope dopants in \rom{2}-\rom{6} 
semiconductors \cite{IttermannHI2000,IttermannPRB1999}. 
Our purpose is threefold: 1) to elucidate the properties of the Li dopant,
2) as a control for future experiments focusing on thin films 
and surface effects in pure and  intentionally doped ZnO, and 3) as a first step 
towards studying more complex related materials, such as indium gallium zinc oxide (IGZO) 
used in thin-film transistors \cite{NomuraNat2004}. In the temperature ($T$) range 
of 7.6 to 400 K, we observe three distinct defects and a broad resonance originating 
from a distribution of metastable complexes. Density functional theory calculations 
of the electric field gradient (EFG) tensor, that determines the quadrupolar splitting 
of the NMR, support the assignment of (diamagnetic) defect configurations for 
the interstitial (\intp{}) and substitutional (\subn{}). The interstitial is ionized and 
stable across the entire $T$ range as a source of self-compensation, while the 
substitutional becomes prevalent only above 300 K as a shallow acceptor. We 
attribute spin-lattice relaxation above 300K to the onset of local defect dynamics. 
We observe a minor fraction of Li in a  third characteristic site and discuss its 
possible origins. Finally, we report the temperature dependence of the EFG 
for the interstitial, which exhibits a remarkable similarity to non-cubic metals. 

\section{Experimental}

\subsection{ Implanted Ion \bnmr } 

In this form of \bnmr, we measure the NMR of a short-lived $\beta$-radioactive probe 
implanted into the crystalline host as a low energy ion beam. Detection uses 
the asymmetric $\beta$-decay (i.e. the direction of the emitted $\beta$ electron 
is correlated to the nuclear spin direction at the instance of the decay), 
similar to \usr \cite{MacFarlaneSSNMR2015}. The experiments were 
performed at TRIUMF's ISAC facility in Vancouver, Canada, where 
a 20-25 keV beam of spin-polarized $^8\ch{Li}^+$ with an intensity 
of $\sim 10^{6}$ ions s$^{-1}$ was focused into a beamspot $\sim2$ mm 
in diameter centred on the sample. In this energy range, SRIM Monte 
Carlo simulations \cite{ZieglerNIMP2010} predict a mean implantation 
depth $\sim 100$ nm (Appendix \ref{appendix_stopping}).
The $^8\ch{Li}$ nuclear spin $\boldmath{I}$ is polarized in-flight using a 
collinear optical pumping scheme with circularly 
polarized light\cite{HatakeyamaPST2002}. The beam is incident upon the 
sample centred in a high homogeneity superconducting solenoid producing 
a field $B_0$ = 6.55 T parallel to the beam and defining the $z$ direction. 
The polarization $p_z$, the expectation value $I_z$ divided by $I$, of 
the implanted \eli\ is monitored through the experimental $\beta$-decay asymmetry, 
\begin{equation}
A(t) = \frac{N_F(t)-N_B(t)}{N_F(t)+N_B(t)} = a_0 p_z(t)
\label{eq:asy}
\end{equation}
\noindent where $N_F$ and $N_B$ are the $\beta$-rates in two 
opposing scintillation detectors downstream (Forward, $F$) and 
upstream (Backward, $B$) of the sample. $a_0$ is a proportionality 
constant depending on properties of both the decay and the detectors. 
The polarization direction is alternated, via the sense of circular polarization 
of the pumping light, between parallel and antiparallel with the beam ($\pm$ helicity), 
and data for each helicity is collected separately to enable accurate baseline 
determination, reduce systematic error, and to sometimes identify helicity dependent features. 
$^8\ch{Li}$ has a nuclear spin $I = 2$, gyromagnetic ratio $\gamma/2\pi = 6.3016$ MHz T$^{-1}$, 
nuclear electric quadrupole moment of $Q = +32.6$ mb\cite{VossJPGNP2011},
and mean radioactive lifetime $\tau = 1.21$ s. The short $\tau$ 
and a typical flux of $\sim 10^6$ ions s$^{-1}$ ensures that the number 
of \eli\ in the sample is never higher than $\sim 10^7$, so the probe 
is ultradilute, and thus Li-Li interactions are absent. 
Further details on the measurements and samples are in section \ref{measurements}.

\subsection{ \eli\ NMR } 

Here we summarize the NMR properties of \eli\ that will 
be important in interpreting the results presented below.

Like its stable isotope counterparts, \eli\ is a quadrupolar nucleus with $I>1/2$, 
meaning its nuclear charge distribution is not spherical, and its nuclear 
spin is thus coupled to the tensor electric field 
gradient (EFG) $V_{ij}$ that it finds at its site in the crystal.
The quadrupolar interaction splits the NMR spectrum into a 
multiplet of $2I=4$ quadrupolar satellites, each corresponding 
to a specific $|\Delta m| = 1$ transition among adjacent nuclear 
magnetic sublevels indexed by $m$. Because $I$ is an 
integer (rare among stable nuclei, but represented,  
for example, by the $I=1$ $^6$Li), there is 
no $m=\pm 1/2$ ``main line'' transition at the centre of the quadrupolar multiplet. 
The scale of the quadrupole splitting is given by the product 
of the EFG with the nuclear electric quadrupole moment $eQ$.
We quantify it by defining the quadrupole frequency by 
\begin{equation}
h \nu_q = \frac{3eQV_{zz}}{4I(2I-1)} = \frac{eQV_{zz}}{8},
\end{equation} 
where $V_{zz}$ is the (largest) principal component of the EFG.
Typical of light nuclei, $Q$ is not very large, and $\nu_q$ is
typically on the kHz scale in close-packed crystals. One can then treat the
quadrupolar interaction accurately as a first order perturbation on the Zeeman Hamiltonian,
whereupon the satellites are distributed symmetrically about the NMR frequency $\nu_r$ 
with positions given by 
\begin{equation}
\nu_i = \nu_r -  n_i \nu_q \frac{1}{2}( \, 3\cos^2\theta-1 + \eta \sin^2\theta\cos2\phi)  
\label{firstordersats}
\end{equation}
\noindent where $n_i$ = $\pm1(\pm3)$\cite{VegaQNS}. $\theta$ and $\phi$ 
are the polar and azimuthal angles of the field $B_0$ in the 
principal axis system of the EFG. The asymmetry parameter $\eta \in [0,1]$ is 
a measure of the deviation of the EFG from axial symmetry ($\eta =0$). If the 
site has threefold or higher symmetry, the EFG will be axial. 
If it is cubic, the EFG (and splitting) are zero.

To measure a resonance spectrum we use a transverse RF field $(B_1)$ that 
induces magnetic dipolar transitions of $|\Delta m| = 1$. When the magnetic field 
of $B_1$ is large enough, we can, in some cases, also observe nonlinear multiquantum 
transitions (MQ), e.g.\ with $|\Delta m| = 2$. These resonances interlace the single quantum 
satellites with positions given by the same formula with $n_i = 0(\pm 2)$. 
Aside from being very strongly dependent on $B_1$, the multiquantum satellites 
are unaffected by a distribution of $\nu_q$ (quadrupolar broadening) resulting 
from crystal imperfections, so they are often noticably narrower 
than their single quantum counterparts\cite{Vega1976PRL}. 
We note that one major source of linewidth in solids, magnetic nuclear dipolar broadening, 
is nearly absent in ZnO, since the nuclear moments of Zn and O are small 
and/or low abundance. We thus expect the predominant broadening will be from crystalline disorder. 

We will use the quadrupolar splitting as a fingerprint of the specific crystallographic 
site for the implanted \eli. Moreover, the EFG can be calculated accurately in 
density functional theory\cite{BlahaPRL1985,SchwarzPRB1990,BlahaPRB1988,DarribaPRB2019}, 
and we performed such calculations
to aid in determining the sites, see section \ref{efgcalc} below.

Aside from the quadrupolar splitting, the NMR frequency $\nu_r$ is shifted from 
the Larmor frequency $\nu_0 = (\gamma/2\pi) B_{0}$ in the applied field by 
the magnetic response of the host. We quantify this shift by comparison 
to a standard reference. With the superconducting magnet in persistence mode, 
our conventional reference is a single crystal of cubic MgO, where the site is cubic, 
and the \eli\ NMR consists of a single narrow line\cite{MacFarlaneJPCS2014}. 
We define the relative shift in parts per million (ppm) as
\begin{equation}
\delta = \frac{\nu_{r}-\nu_{\ch{MgO}}}{\nu_{\ch{MgO}}} \cdot 10^6.
\label{relshift}
\end{equation}

The resonance shift has contributions from the macroscopic demagnetization field, 
local magnetic fields of unpaired spins in the vicinity, and the 
local screening response, i.e.\ the chemical (orbital) shift \cite{MacFarlaneSSNMR2015}. 
The latter is well-known from stable Li NMR to exhibit a very small range of a few ppm; 
while the demagnetization field depends on the shape of the sample, and it is near 
its largest in our geometry: a thin plate in a perpendicular field.

In addition to the time average local fields that determine the spectral features, 
fluctuations of these fields at the NMR frequency determine the spin-lattice relaxation (SLR).
Like conventional NMR, \eli\ relaxation is described by a relaxation 
function (recovery curve) typically parametrized by the rate $1/T_1$. 
The finite \eli\ lifetime $\tau$ limits measurable $T_1$ roughly to the 
range $0.01\tau < T_1 < 100\tau$. Because the spin is polarized in flight, 
no RF is needed to measure the relaxation, but as a result, there is no 
spectral resolution to the relaxation which is simply the average of all 
the \eli\ in the sample. The type of fluctuations giving rise to the observed 
SLR can often be determined by the characteristic temperature dependence 
of $1/T_1$. In a nonmagnetic insulating host, fluctuations may arise from 
stochastic diffusive motion of interstitial \eli\ or from lattice vibrations (phonons). 
In a semiconductor, mobile or localized carriers, if they are sufficiently abundant, 
may provide other sources for relaxation \cite{Selbach1979}.

\subsection{ \bnmr\ Measurements }
\label{measurements}

We used two transparent colorless commercial hydrothermally grown 
ZnO single crystals (Tokyo Denpa Co.\ Ltd., Tokyo) in the form of thin 
plates $10\times8\times0.5$ $\mathrm{mm}^3$ perpendicular to (0001), 
the hexagonal crystallographic $c$ axis \cite{MaedaSST2005}. One sample 
was used ``as-grown" (AG) while the other(n$^+$) was annealed at \SI{1400}{\degreeCelsius} to 
remove impurities. Typical AG carrier concentrations around room temperature 
are $10^{12}$ - $10^{15}$ cm$^{-3}$, while annealing raises it 
to $10^{16}$ - $10^{17}$ cm$^{-3}$ by eliminating compensating 
defects (such as Li) residual from the growth process. 
Annealing was done prior to polishing and the polished surfaces 
had step-and-terrace structure with a typical step height $\sim 0.5-3$ nm. 
The same crystals were used for \usr\ experiments which showed a 
shallow muonium signal, typical of high quality ZnO. For the measurements, the crystals were 
clamped to the Al sample holder of an ultra-high vacuum cold finger He flow cryostat. 

With $B_0$ = 6.55 T parallel to the crystalline $c$-axis, we performed three 
types of measurement as a function of temperature. 
A) With a continuous beam, we measured the resonance spectrum by 
stepping the frequency of a continuous wave transverse RF magnetic 
field $B_1$ through the Larmor frequency. On-resonance, the RF 
causes the \eli\ spin to precess, decreasing the time integrated asymmetry.

B) A second type of resonance measurement was used in which 
the RF consisted of a four frequency comb
with equal amplitude oscillations at frequencies determined by two 
parameters: $\tilde{\nu}_0$ and $\tilde{\nu}_q$. Specifically, at $\tilde{\nu}_0 \pm 3\tilde{\nu}_q$ 
and $\tilde{\nu}_0 \pm \tilde{\nu}_q$. The center $\tilde{\nu}_0$ was fixed 
at the resonance frequency from the normal spectrum described in A above, 
and the splitting parameter $\tilde{\nu}_q$ was then stepped through a range of values.
The quadrupole satellites in the single mode RF spectrum suffer from small amplitudes, 
as a single transition is limited to at most 25\% of the polarization. As a result, only a 
few spectra were obtained, which take considerable time to accumulate. In contrast, 
with the RF comb, when $\tilde{\nu}_q$ matches the quadrupole splitting, all the 
single quantum transitions ($\Delta m = 1$) can be saturated at once, dramatically 
increasing the amplitude, even revealing otherwise undetectable signals. The RF comb 
method has been previously applied to $^{12}$B and $^{21}$Na \bnmr\ in \rom{2}-\rom{6} semiconductors 
and a more detailed description can be found elsewhere \cite{MinamisonoNIMPR2010,IttermannPRB1999,Minamisono1993}. 

Both types of resonance measurement differ in an important way from familiar conventional 
pulsed NMR, because the RF is applied continuously at each step 
of the scan for a relatively long integration time (1 s).
In this mode, the resonance amplitudes are determined by several factors 
beyond the relative abundance of the corresponding \eli\ site. These include the RF 
amplitude $B_1$, the resonance linewidth, and, more subtly, slow dynamics of the 
resonance frequency on the integration timescale which can cause \eli\ spins to 
be double counted, since an individual spin need only be on-resonance for a 
millisecond or so during the integration time to be fully depolarized. As a result, 
it is difficult to make quantitative conclusions about site fractions from the amplitudes.

C) Measurements of the spin-lattice relaxation used a pulsed \elip\ beam and no RF field. 
The time evolution of the polarization was measured both during and after the 4 second pulse. 
At its trailing edge, the polarization approaches a dynamic steady state, while after the pulse, 
it decays to its thermal equilibrium value near zero, giving rise to the characteristic bipartite 
NMR recovery curve. The helicity was alternated for each pulse every 20 seconds, and 
this process was repeated for about 30 minutes to accumulate statistics.

\subsection{ EFG Calculations }
\label{efgcalc}

To facilitate site assignment for the implanted \eli, the EFG was 
calculated for various Li defect configurations using the supercell method. 
Here, Li is placed at a specific site in a supercell composed of a small 
number of ZnO unit cells which is then subjected to periodic boundary conditions. 
The calculation thus represents a fictitious ordered phase Li$_x$ZnO. 
While $x$ in the calculation is vastly larger than in the experiment, by 
decreasing $x$ to the extent possible, one can approach the dilute limit, 
particularly for quantities like the EFG that are predominantly sensitive to the 
immediate environment \cite{LanyPRB2000, WichertHI2001,BlahaPRB1988}.

Structural calculations of the defects were performed within the density functional 
theory framework using the plane wave pseudopotential method as implemented in 
Quantum Espresso \cite{GiannozziJPCM2017}. Troullier-Martins norm-conserving 
pseudopotentials \cite{TroullierPRB1991} were employed for treatment of core electrons 
and the generalized gradient approximation PBE was used for the exchange 
and correlation functional \cite{PerdewPRB1996}. A supercell of $5a\times5b\times3c$ lattice 
units was used with a kinetic energy cutoff of 1080 eV and $2\times2\times2$ Monkhorst-Pack 
grid (MPG) \cite{MonkhorstPRB1976} centred at the $\Gamma$ point. The supercell 
had \ch{Li} defects introduced at various sites with a compensating background jellium 
charge to maintain overall neutrality and avoid divergent Coulomb interactions. 
The atomic positions were allowed to fully relax under the constraint of fixed experimental 
lattice parameters \cite{ReeberJAP1970}. The total energy was numerically converged to 
less than 1 meV with respect to a cutoff of 1210 eV and $6\times6\times6$ MPG. 

For an isolated implanted \lip, the surrounding ZnO lattice will respond with a 
localized distortion about the defect that leaves the more distant structure unchanged. 
The EFG has an $r^{-3}$ dependence on distance and is thus very sensitive to the local 
structure, so a realistic calculation requires both the experimental lattice parameters 
and an accurate estimate of the local lattice relaxation \cite{BlahaPRB1988}.

EFG calculations were performed using the more computationally expensive 
all-electron augmented plane wave plus local orbitals method (APW+lo) 
implemented in WIEN2k\cite{BlahaJCP2020} and the relaxed structure from the 
pseudopotential method. The angular momentum expansion of the lattice harmonics 
inside the atomic spheres was truncated at $L=10$ and the plane waves outside the 
atomic spheres expanded with a cutoff of R$_{\mathrm{MT}}\mathrm{K}_{\mathrm{Max}} = 8$, 
the product of the smallest atomic sphere (R$_{\mathrm{MT}}$) with the largest 
K-vector ($\mathrm{K}_{\mathrm{Max}}$) of the plane wave expansion. 

\section{Results and Analysis} 

\subsection{Resonance Spectra}

\begin{figure}[ht]
\begin{center}
\includegraphics[width=\columnwidth]{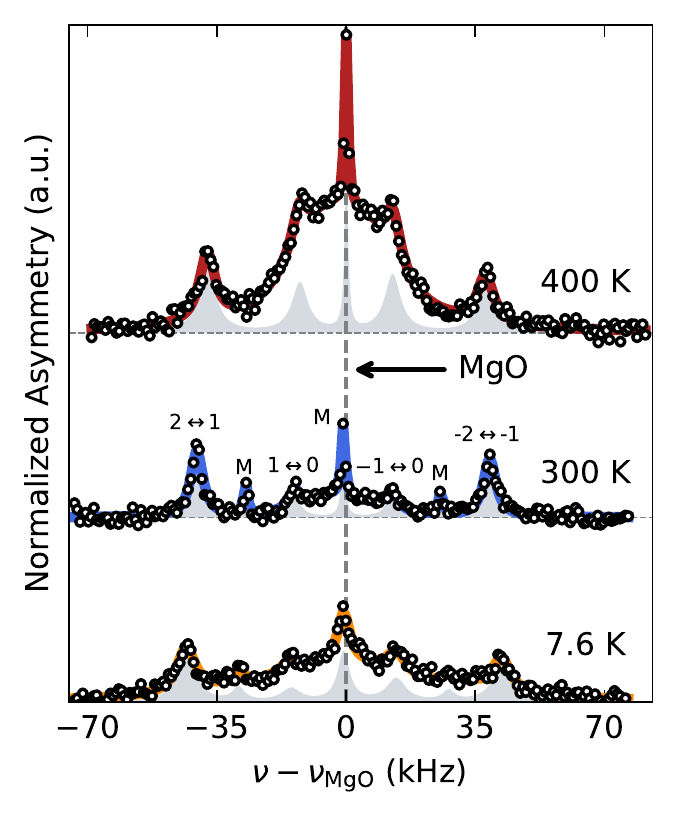}
\caption{The quadrupole split single tone RF spectra at several temperatures 
for \eli\ in the as-grown ZnO single crystal with 6.55 T$\parallel c$. 
The frequency scale is relative to the MgO calibration frequency (dashed line). 
The vertical scale has been normalized by
the off-resonance asymmetry at each temperature, and the spectra are vertically offset for clarity.
At 300 K, the single quantum satellites are labelled with their corresponding $m$ values, 
and multiquantum with M. The fitted A site spectrum is shaded grey 
and is the dominant feature. Above 300 K, a second unresolved line (site B) emerges. }
\label{figure:300K_AG} 
\end{center}
\end{figure}

\Fig{\ref{figure:300K_AG}} shows the single tone frequency spectrum in 
ZnO at three temperatures. At all temperatures, the primary feature is a 
quadrupolar multiplet pattern spread $\pm 45$ kHz about its center. 
At 300 K, the predominant quadrupolar splitting is indexed by the corresponding 
magnetic sub-level transition. In addition to the indexed single quantum 
satellites (SQ), there are interlaced multiquantum lines labeled M 
that are substantially narrower than their SQ counterparts. 

The quadrupole splitting is expected, since no site in the wurtzite lattice is cubic.
However, its magnitude (e.g. defined by the difference between the outermost satellites) 
is quite modest, $\sim 10\times$ smaller than in perovskite oxides\cite{MacFarlanePB2003}. 
The resolved multiplet indicates a well-defined crystalline site, which 
we label ``A''. The A multiplet is conspicuous at all temperatures, 
but its features evolve, with increased splitting and substantial broadening at low $T$.
At the lowest temperature, the multiplet is superimposed on a broad pedestal 
of intensity with a width comparable to the overall multiplet splitting.
In addition to A, at 300 K and above, there is another unresolved line close to the Larmor frequency 
(see the 400 K spectrum in \Fig{\ref{figure:300K_AG}}). The helicity-resolved spectra 
confirm it to be quadrupolar but with a substantially smaller splitting, 
such that the individual satellites are not resolved (see Appendix \ref{appendix_resonance}).
We conclude this corresponds to a second \elip\ site (B), with a much smaller EFG,
whose population grows at high temperature.

\begin{table*}[t]
\centering
\caption{The shared best fit parameters for the A multiplet, including values for the 
quadrupole frequency $\nu_q$, frequency shift $\delta$, single quantum linewidths $\Delta \nu_S$, 
and multiquantum linewidths $\Delta \nu_M$ for both AG and \np\ samples. 
The errors in parentheses are purely statistical. }
\renewcommand*{\arraystretch}{1.4}
\begin{tabular}{ c  c  c  c  c  c  }
\hline
 Sample& Temperature (K) & $\nu_q$(kHz) & $\delta$(ppm) & $\Delta \nu_S$(kHz)  & $\Delta \nu_M$(kHz)  \\
 \hline
 \hline
AG &7.6 & $14.23(2)$ & $9.3(1)$ & $6.0(4)$ & $4(1)$       \\ 
AG&300  & $13.22(1)$ & $5.9(7)$ & $3.9(3)$  & $1.2(2)$     \\ 
AG&400  & $12.56(2)$ & $1.7(5)$ & $5.2(2)$  & $1.08(5)$    \\ 
n$^+$&300  & $13.16(1)$ & $1.0(6)$ & $3.9(1)$  & $1.38(6)$    \\ 
\hline
\end{tabular}
\label{table:linewidth}
\end{table*}

To quantify these observations, the data was fit using using a custom python 
code based on the Minuit2 library \cite{JamesCPC1975,Hatlo2005}. 
At each temperature, the two helicities were fit simultaneously sharing the parameters 
that determine the satellite positions. Note that for A, $\nu_q/\nu_r \sim 3\times10^{-4}$, 
so Eq.\ (\ref{firstordersats}) provides the satellite positions very accurately. 
The fit function thus consisted of 7 Lorentzians with satellite positions determined by 
the single parameter $\nu_q$ via Eq.\ (\ref{firstordersats}), assuming the $c$-axis 
coincides with the principal EFG direction ($\theta =0$) and axial symmetry ($\eta =0$). 
We discuss the validity of these assumptions below. It was further constrained by assuming 
all the SQ (MQ) satellites shared the same width $\Delta \nu_S$ ($\Delta \nu_M$) at each temperature.
Where necessary, signals corresponding to the low $T$ pedestal and the unresolved B site signal were added. 
{\it A priori} we expect the chemical shift at A and B will differ, 
but any such difference is too small to detect, and to reduce the number of parameters, the 
broad lines were centred at the same $\nu_r$ as A.
The resulting fits are shown as the lines in \Fig{\ref{figure:300K_AG}}, 
and the shared parameters for A are given in \Table{\ref{table:linewidth}}.

The resulting $\nu_q$, determined by the splittings, are quite precise, and the differences
in \Table{\ref{table:linewidth}} are significant as confirmed below by the comb spectra.
We include the raw relative shift defined by Eq.\ (\ref{relshift}) in \Table{\ref{table:linewidth}}.
From the sample shape, we estimate the dimensionless demagnetization factor is $N \approx 0.92$.
A literature \cite{MikhailJPCS1966} value for the volume susceptibility of 
ZnO $\chi_v \approx -2.2\times 10^{-6}$ leads to an estimated demagnetization shift
$-4\pi \left( N-1/3 \right) \chi_v \approx 16$~ppm,
comparable to the value in the MgO reference crystal ($\sim 12$ ppm\cite{MacFarlaneJPCS2014}).
From this, we conclude an accurate estimate of the corrected shift is impractical.
However, the chemical shifts in MgO and ZnO are evidently small (a few ppm) and similar, 
consistent with expectations from conventional Li NMR. 
The reported shift uncertainties in \Table{\ref{table:linewidth}} are purely statistical. Unlike the splitting,
we do not regard the differences in the shift as significant, because it is sensitive to
small systematic changes in the field at the sample, due to thermal contraction, 
weak magnetization of the spectrometer, and reproducibility of the sample position.

The annealed sample was measured briefly at 300 K (not shown). 
The SQ and MQ satellite linewidths at 300 K in the two samples are similar, while any difference in the 
quadrupole splitting is much smaller than the MQ linewidths and indistinguishable in a single tone spectrum. 

In the AG sample, both the SQ and MQ satellites broaden significantly, 
and by a similar amount, at low temperature, indicating predominantly magnetic, 
rather than quadrupolar, broadening.
Interestingly, the SQ satellites also broaden at high temperature above 300 K. 
The $n_i = \pm 2$ MQ satellites were not observed at 400 K, so no comparison of their widths can be made.

Aside from the A multiplet, the broad lines are considerably less well-determined. 
The low temperature pedestal is very similar in the two helicities, 
meaning it shows little evidence of a resolved quadrupole splitting - or any such splitting 
is much less than the width which is on the order of $50$ kHz FWHM. 
Without resolved satellites, it is not possible to reliably extract the quadrupolar splitting of the B site. 
From the helicity resolved fits, 
we estimate a splitting of at most a few kHz with satellite linewidths several times larger than this.

\subsection{Comb Spectra}

\begin{figure}[th]
\begin{center}
\includegraphics[width=\columnwidth]{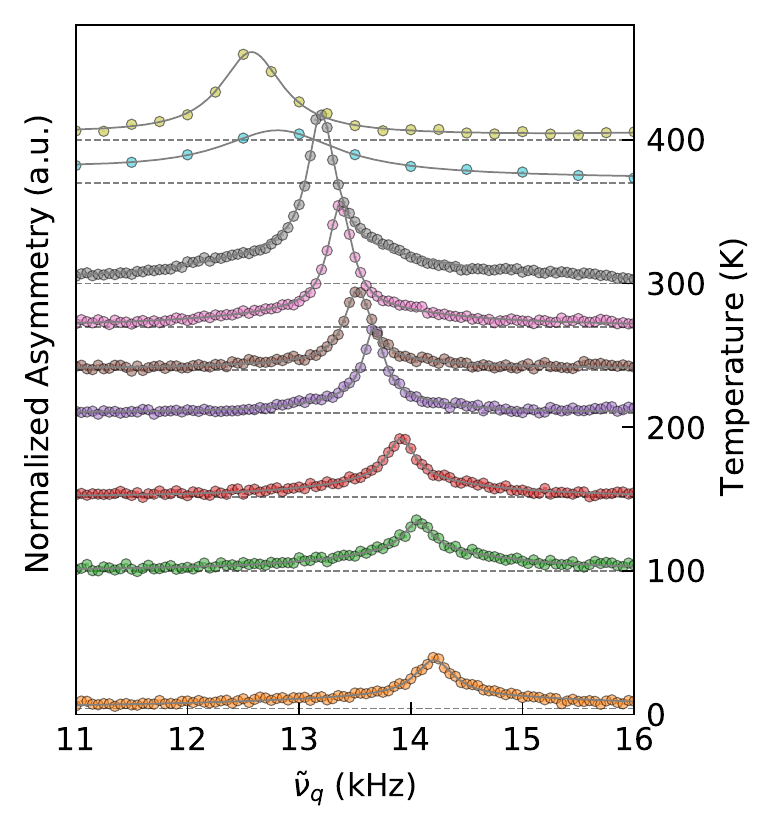}
\caption{The RF comb spectra, for a narrow range around $\nu_q$ for site A, as a function of temperature. The spectra are vertically offset by an amount proportional to the temperature. There is a clear reduction of $\nu_q$ with increasing T. The solid grey lines are bi-Lorentzian fits. Up to 300 K, the amplitude increases. Above this, there is a dramatic decrease and extensive broadening.} 
\label{figure:comb_data} 
\end{center}
\end{figure}

We turn now to the RF comb spectra that reveal more detail than the single frequency spectra above. 
Recall the central frequency of the comb is set to coincide with the centre of the multiplets 
in \Fig{\ref{figure:300K_AG}} and then the comb splitting is stepped over a range. 
When the comb's splitting parameter $\tilde{\nu}_q$ matches $\nu_q$, we find a 
single resonance much larger than any of the individual satellites. 
This is demonstrated for the $A$ site multiplet as a function of temperature in \Fig{\ref{figure:comb_data}}. 
The resonant $\tilde{\nu}_q$ is entirely consistent with the splitting in \Fig{\ref{figure:300K_AG}}. 
With increasing temperature, the resonance position moves systematically downward, 
confirming the trend in \Table{\ref{table:linewidth}} and revealing a thermal reduction in 
EFG which we discuss below in section \ref{EFGT}. The spectra could not be fit with a single Lorentzian, 
but required a bi-Lorentzian (narrow plus broad) sharing the same $\nu_q$. The fit parameters, 
as a function of temperature, are shown as the triangles in \Fig{\ref{figure:comb}}. 
The resonance position provides a very accurate measurement of $\nu_q(T)$. 
The amplitude increases with increasing $T$, reaching a maximum at $\sim 300$~K, 
while it is widest at 370 K, consistent with the high temperature broadening of the SQ satellites in \Table{\ref{table:linewidth}}.

\begin{figure}[th]
\begin{center}
\includegraphics[width=\columnwidth]{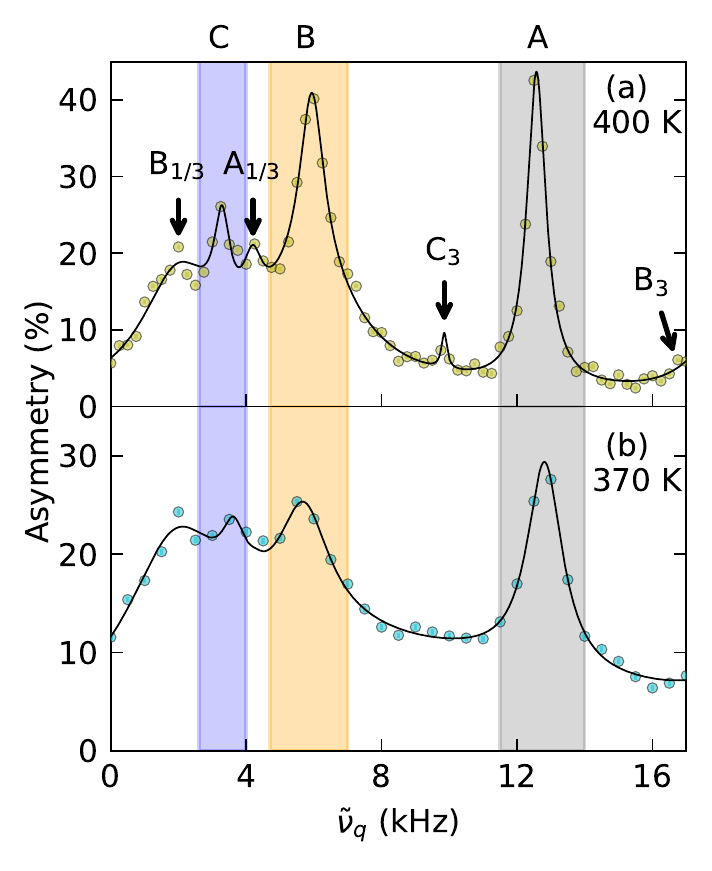}
\caption{The RF comb spectra for a broader range down to $\tilde{\nu}_q=0$ at 400 K (a) and 370 K (b). 
Primary resonances for distinct sites are indicated with shaded regions, labelled A,B,C. 
The asymmetry has been normalized to its fitted off-resonance value. 
At 370 K, the lines are all broadened. Smaller features denoted as subscripted labels are aliases of the primary resonances. 
There is also a substantial broad background.
} 
\label{figure:comb_full_spectrum} 
\end{center}
\end{figure}

For only a few high temperatures, we extended the $\tilde{\nu}_q$ scan range down to zero comb splitting. 
These spectra, shown in \Fig{\ref{figure:comb_full_spectrum}}, reveal further structure. 
Distinct well defined quadrupole splittings for $\tilde{\nu}_q = \nu_q$ resonances are denoted 
by the vertical colored bands, with the grey band corresponding to the A 
resonance tracked with $T$ in \Fig{\ref{figure:comb_data}}. In scanning $\tilde{\nu}_q$ over a wider range, 
it is important to recognize that when the outer(inner) comb frequencies match the inner(outer) SQ satellites of 
the underlying multiplet, we expect small aliased resonances at $\tilde{\nu}_q = \nu_q/3$ (and $3\nu_q$). 
These SQ resonances are denoted at 400 K by subscripts. 
There may also be aliased MQ resonances at $2\nu_q/3$ and $2\nu_q$. 
Scans of $\tilde{\nu}_q$ up to 82, 110 kHz at 300 and 400 K (not shown) reveal 
no further resonances with larger $\nu_q$. The features at lower frequency includes a 
resolved resonance corresponding to the B site (shaded orange), with $\nu_q = 5.941(4)$ kHz at 400 K, 
and fit values as open squares in \Fig{\ref{figure:comb}}. 
This resonance is also identifiable at 210 K (not shown), indicating a small population of 
site B far below where it becomes evident in the single frequency spectra. 
The B resonance is also broader than A, and its position shifts higher from 370 to 400 K opposite to the reduction shown by A.

As discussed in section \ref{measurements}, the resonance amplitudes are determined by several factors, 
including the long integration time (1 s) of the continuously applied RF field and off-resonance baseline asymmetry. 
In the presence of slow spectral dynamics up to the integration time, 
the amplitude may be enhanced as \eli\ can be double counted if they 
only transiently meet the resonance condition for a time as short as the RF precession period, 
on the order of 1 ms. This results in spectra with resonance amplitudes whose sum 
may exceed the full off-resonance asymmetry, as is evidently the case in 
\Fig{\ref{figure:comb_full_spectrum}}. This is strong confirmation that \eli\ is undergoing spectral dynamics in this temperature range.

The comb spectrum also reveals a third site (labelled C) with an even smaller
\nuq{} = 3.63(5) kHz at 370 K, about half the value for B, decreasing to 3.28(5) kHz at 400 K.
It's amplitude is small, so it corresponds to only a minor population of implanted \eli. 
The combination of the breadth of B and the presence of C explains the 
absence of resolved structure of the B line in \Fig{\ref{figure:300K_AG}}, but note 
there is substantial intensity in a broad background encompassing B and C with a tail that
reaches toward A. The three resolved quadrupole frequencies at 400 K are given in \Table{\ref{table:sitecomp}} 
and the additional minor peaks are aliases of these resonances.
The rich detail in these spectra demonstrates the power of the comb method to strongly 
amplify quadrupolar split resonances within a narrow band of $\tilde{\nu}_q$ and reveal structure that is otherwise hidden.

\begin{table}[h]
\centering
\caption[]{Measured quadrupole frequency $\nu_{q}$ and principal component of the EFG 
tensor $V_{zz}$ for the three \eli\ sites at 400 K. Parenthetical values are statistical errors.}
\label{table:sitecomp}
\renewcommand*{\arraystretch}{1.4}
\begin{tabular}{ c c c c c c } 
\hline
 Site & $\nu_q$(kHz) & $V_{zz}$($10^{20}$V/m$^2$)  \\
 \hline
 \hline
 A & $12.56(2)$ & $1.275(2)$ \\
 B & $5.941(4)$ & $0.6029(4)$ \\
 C & $3.28(5)$ & $0.333(5)$ \\
\hline
\end{tabular}
\end{table}

\begin{figure}[ht]
\begin{center}
\includegraphics[width=\columnwidth]{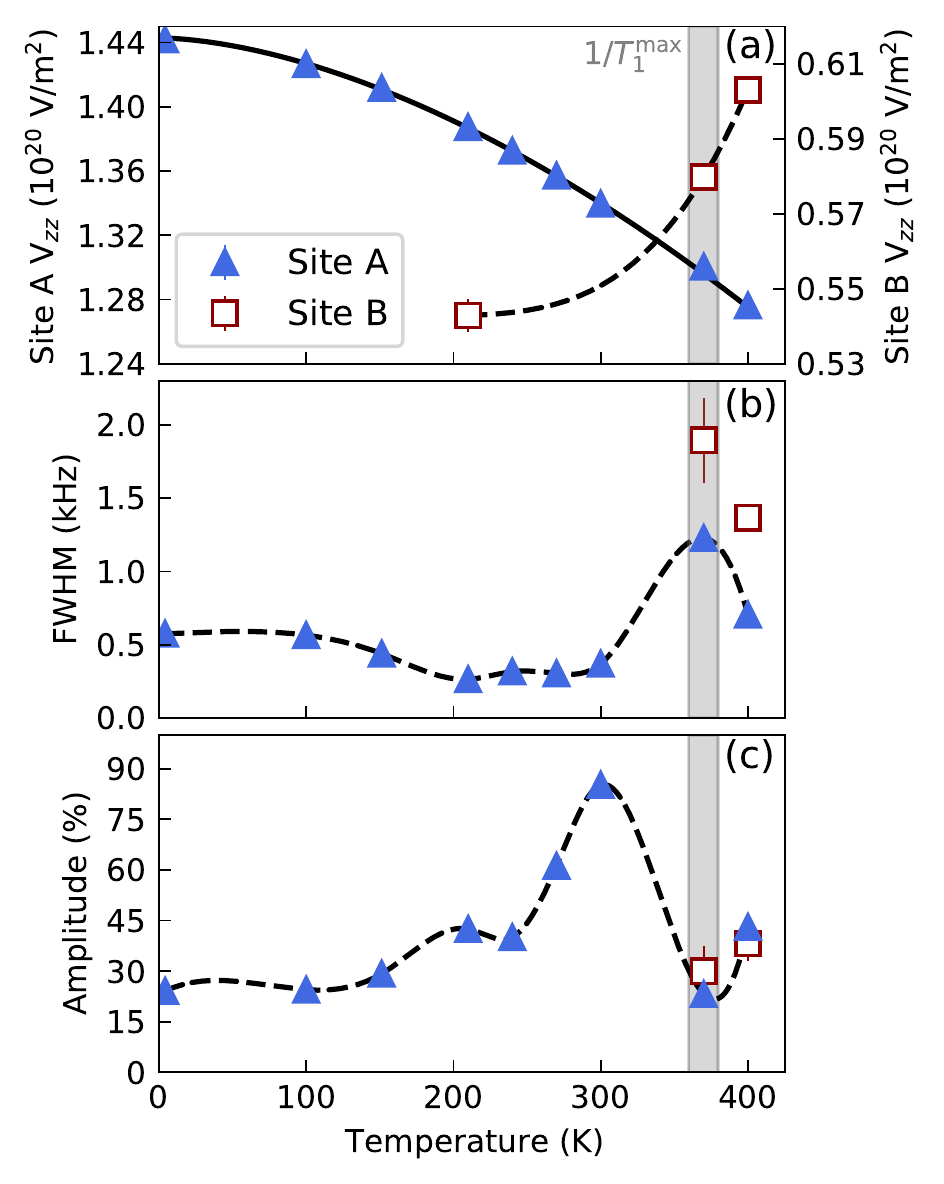}
\caption{Temperature dependence of the fit parameters for the comb spectra in 
Figs.\ \ref{figure:comb_data}, \ref{figure:comb_full_spectrum}.
(a) From the resonance positions, the principal component of the EFG for sites A and B calculated using Eq.\ \ref{firstordersats}.
(b) the linewidth (FWHM) for the bi-Lorentzian. (c) the resonance amplitude.
The lines are guides to the eye, except the solid line in (a) is a fit to Eq.\ \ref{eq:EFG}
for site A. The vertical band marks the position of
the peak of the spin-lattice relaxation rate.}
\label{figure:comb} 
\end{center}
\end{figure}

\subsection{Spin-Lattice Relaxation}

\begin{figure}[ht]
\begin{center}
\includegraphics[width=\columnwidth]{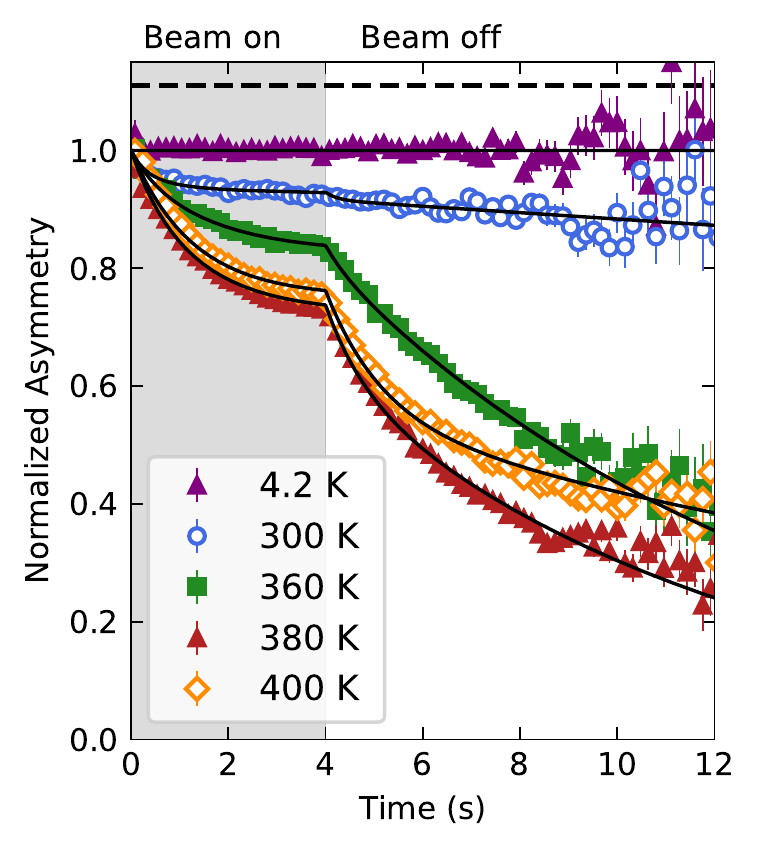}
\caption{The time dependence of the asymmetry for \elip\ in the as-grown ZnO at 6.55 Tesla$\parallel c$ for several temperatures. 
The decay is due to spin-lattice relaxation of the isolated implanted \elip. 
The black lines are global fits with a biexponential relaxation function Eq.\ (\ref{eq:relaxation_fcn}). 
Above 300 K, there is a strong temperature dependence with a maximum rate near 370 K.
The dashed line shows the calibrated value of $A_0$ in MgO exceeds that in ZnO indicating a small missing fraction.}
\label{figure:SLR} 
\end{center}
\end{figure}

\begin{figure}[ht]
\begin{center}
\includegraphics[width=\columnwidth]{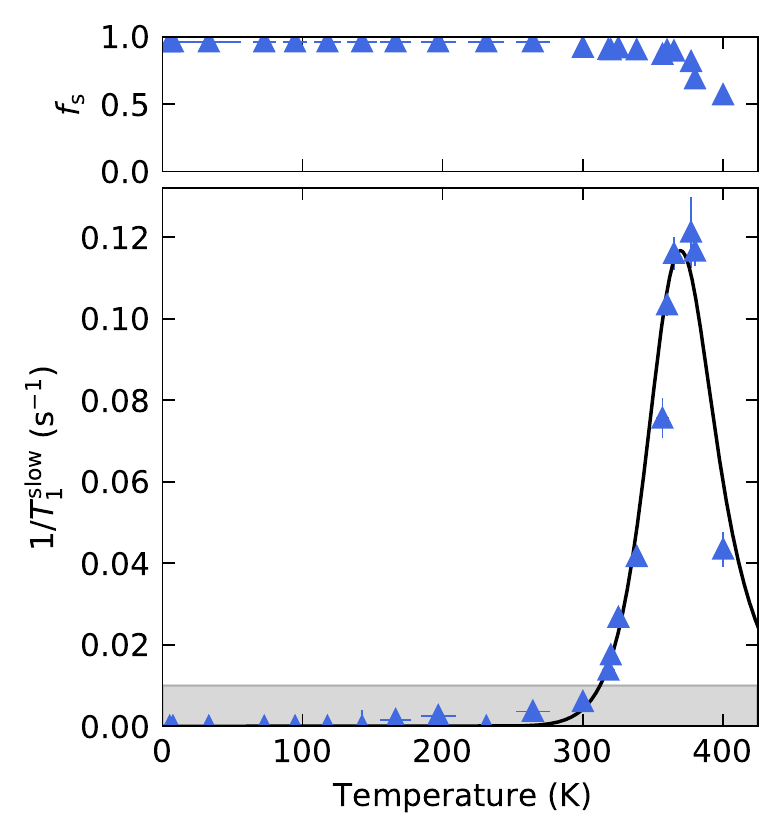}
\caption{The spin-lattice relaxation rate, $\lambda_s = 1/T_1^{\mathrm{slow}}$ 
and fraction $f_s$ for the large slow relaxing component from biexponential fits as 
shown in Fig.\ \ref{figure:SLR}. Near 400 K, $f_{s}$ decreases as more of the 
signal is accounted for by the fast component. There is a clear maximum
at 370 K. The line is a fit to an activated BPP temperature dependence as described in the text. The grey region denotes the lower limit at which $1/T_1$ is accurately measured.}
\label{figure:SLR_rates} 
\end{center}
\end{figure}

Below 300 K, the spin-lattice relaxation in the AG sample is very slow ($T_{1} > 100 \ \mathrm{s}$), 
as seen in the recovery curves in \Fig{\ref{figure:SLR}}, where the time dependent 
asymmetry has been normalized to its initial value $A_0$. 
The corresponding rate is near the limit imposed by the \eli\ lifetime. 
This is typical for a nonmagnetic insulator when \elip\ is not diffusing. However, above 300 K it increases rapidly, 
but not monotonically, with a maximum rate at about 370 K (the lowest data in \Fig{\ref{figure:SLR_rates}}). 
Careful comparison between 300 K and low temperature shows there is a small much faster relaxing component, 
corresponding to at most a few \% of the signal. In addition, a calibration of the initial asymmetry $A_0$ in MgO
is about 10\% larger at 300 K, see the dashed horizontal line \Fig{\ref{figure:SLR}}. 
Thus, there appears to be a small ``missing fraction'' corresponding to a population of \eli\ that are very rapidly depolarized. 

We briefly studied the \np\ sample at 300 K and found the SLR rates to be similar. 
However, $A_0$ was larger, only 3.3(1)\% less than MgO (see \Fig{\ref{figure:SLR_annealed}} in Appendix \ref{appendix_SLR}), 
i.e.\ the missing fraction is substantially reduced by annealing.

To fit the data consistently across the full temperature range, 
we adopt a biexponential relaxation function: for \eli\ arriving at time $t^\prime$, 
the polarization at time $t>t^\prime$ follows 
\begin{equation}
R(t,t^\prime) = [ f_s e^{-\lambda_s(t-t^\prime)}  + (1-f_s)e^{-\lambda_f(t-t^\prime)} ] 
\label{eq:relaxation_fcn}
\end{equation}
\noindent where $f_s$ is the slow fraction and $\lambda_s = 1/T_1^{\mathrm{slow}}$ is its rate, 
while $\lambda_f$ is the fast SLR rate. The data was fit to Eq.\ (\ref{eq:relaxation_fcn}) convolved 
with the 4 second beam pulse using the Minuit2 \cite{JamesCPC1975,Hatlo2005} library 
as implemented in ROOT\cite{BrunNIMPRA1997} with a custom C++ code.
To reduce the number of free parameters, below 300 K $A_0= 0.07822(5)$ and $f_s = 0.962(1)$ were 
shared temperature independent global parameters, and the global reduced $\chi^2 \sim 0.85$. 
The data above 300 K was taken after the spectrometer had been modified to increase its maximum temperature. 
For this data, $f_s$ was allowed to vary with temperature and the shared $A_0 = 0.0854(2)$ with $\chi^2 = 0.87$. 
This increase in $A_0$ is due to changes to the F $\beta$-detector. 
In both ranges, the $\chi^2$ values indicate the biexponential tends to overparametrize the data (particularly at low temperature); 
however, it has the advantage that it isolates the fast component, preventing it from biasing the slow rate. 
The rate and fraction of the large, slow-relaxing signal are presented in Fig.\ \ref{figure:SLR_rates} showing 
the rapid increase in $\lambda_s$ above 250 K toward a maximum at $370\pm10$ K. 
The fast signal is a very small fraction below 300 K, see Appendix \ref{fastcomp} for more details.

\subsection{EFG calculations} 

As the EFG is a ground state property of the charge density which is accessible within 
all-electron self-consistent energy band calculations \cite{BlahaPRL1985}, we attempted to determine 
the site of \eli{} by comparison with the calculated EFG tensor magnitude and symmetry. 
In semiconductors, defects generally exhibit distinct charge states, and the lowest energy charge state may change 
with temperature (or doping) by exchanging electrons with band states at the Fermi level. 
Naturally, different charge states have different local lattice and electronic structures\cite{FreysoldtRMP2014}. 
This modifies the EFG, allowing, in principle, identification of charge states, particularly 
for deep level defects characterized by large lattice relaxation at a charge state transition \cite{LanyPRB2000, WichertHI2001}. 
We indicate the charge state of the calculation with a superscript $q$,
where $q = -1 (+1)$ if an electron is added to (removed from) the supercell.

\begin{table}[h]
\centering
\caption[]{ The electric field gradient $V_{zz}$ in units of $10^{20} \ \mathrm{V} /\mathrm{m}^2$ for 
selected relaxed {Li} defect configurations using the 300 K ZnO lattice parameters.}
\label{table:calculated_EFG}
\renewcommand*{\arraystretch}{1.4}
\begin{tabular}{ c c c c c c } 
\hline
Site &  $V_{zz}$ & Principal Axes & $\eta$ \\ 
\hline
\hline
\intp{} & 1.21 &(0,0,1) &  0 \\ 
Li$_\mathrm{T}^{\star +}$& 0.47 & (0.04,0.82,-0.57) &  0.65 \\ 
\subn{} & 0.16   & (0,0,1)  &  0\\ 
\hline
\end{tabular}
\end{table}

\begin{figure}[ht]
\begin{center}
\includegraphics[width=\columnwidth]{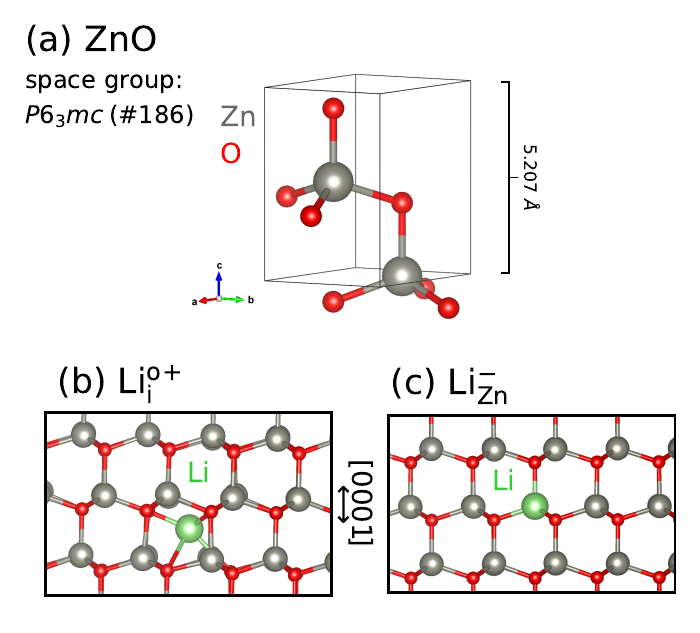}
\caption{Unit cell of ZnO (a) and the relaxed defects of \intp{} (b) and \subn{} (c) in 
their diamagnetic states shown $\| c$ and $\perp b$. The off-center \intp{} is located at the $2a$ site with 
fractional coordinates (0,0,0.079) of the unit cell shown. The structures were drawn using {VESTA} \cite{MommaJAC2011}. } 
\label{figure:crystal_structure} 
\end{center}
\end{figure}

There are only a few reasonable candidate Li sites in ideal ZnO, with structure shown in \Fig{\ref{figure:crystal_structure}}.
We calculate the defect structures for \lip\ in the high symmetry tetrahedral \Tint\ (Wykoff $2b$) and 
octahedral \inth\ (Wykoff $2a$) interstitial sites, where it would act as an electronic donor. 
We also calculate the zinc substitutional site (Li$_{\mathrm{Zn}}$), an established 
acceptor \cite{OrlinskiiPRL2004}. Note that neutral Li$_{\mathrm{Zn}}$ has a polaronic 
distortion around the localized hole, so it is not certain whether it is a 
shallow (and potentially useful) $p$-type dopant \cite{CarvalhoPRB2009,VidyaJAP2012}. 
The Li antisite defect (Li$_{\mathrm{O}}$) is not considered, as it is reported to be unstable \cite{VidyaJAP2012}. 

We find \intp{} is displaced from the symmetric $2a$ position along (0001) 
by $0.57$ \AA, creating a long and short set of Li-O coordinating bonds. 
The axially symmetric \Tp{} is also found to relax spontaneously to Li$_\mathrm{T}^{\star+}$, a 
variant displaced from the 3-fold axis of symmetry. Considering the defects only in their 
diamagnetic ionized states, the EFG tensors, calculated using the 300 K lattice parameters 
are given in \Table{\ref{table:calculated_EFG}}. The \subn\ center distorts the local environment, 
giving rise to an almost symmetric tetrahedral coordination with neighboring oxygens, 
resulting in a very small V$_{zz}$. \intp\ retains axial symmetry (along $c$), while Li$_\mathrm{T}^{\star+}$ has a 
principal axis almost orthogonal to $c$, nearly parallel to $b$. 
The calculated structures of the \intp{} and \subn{} defects are shown in \Fig{\ref{figure:crystal_structure}}. 

\section{Discussion} 
\label{section:discussion}

\subsection{Sites for Implanted Li$^+$}

When an ion is implanted in a crystal, it loses energy through collisions with host lattice atoms and gradually slows down. 
In the process, it can knock a host atom out of its normal lattice position into a nearby vacant interstitial site, 
forming a Frenkel (vacancy-interstitial) pair \cite{Rimini1995}. The threshold for this process is typically 10s of eV. So, 
after generating its final Frenkel pair, the implanted ion continues for some distance before stopping typically at a high symmetry site. 
The host lattice will then relax around the stopped implant, possibly lowering its site symmetry. 
Now if we consider the NMR spectrum, such as we measure with \bnmr, to the extent that the stopping site is isolated from other defects, 
it will have a characteristic EFG tensor reflected by a well-resolved quadrupolar splitting. 
However, it is also possible that the ion stops close enough to the last Frenkel pair that its EFG is modified. 
As there are many possible configurations of the trio (ion+pair), this results in a distribution of EFGs and is 
reflected in a quadrupolar broadened resonance usually observed at low temperature. 
At higher $T$, the Frenkel defects tend to heal rapidly, typically leading to the first stage of annealing. 
Depending on relative mobilities, the implanted ion may compete for the vacancy with the intrinsic interstitial of the pair.
With increasing $T$, the typical behaviour is thus a gradual loss of the broad ``perturbed site'' resonance together 
with an increase and sharpening of the well-defined lines corresponding to the isolated site, 
and potentially (if the implant reaches the vacancy) a change of site to substitutional. 
As we shall see below, the implanted \elip\ in ZnO follows this phenomenology quite closely.

The A site multiplet, with a well-resolved quadrupolar splitting, is evident at all temperatures, 
including the lowest, see Fig.\ \ref{figure:300K_AG}, so it should correspond to \elip\ at the most stable interstitial. 
The octahedral site is substantially more spacious than the tetrahedral, and
theory finds it lower in energy by 0.62 eV\cite{CarvalhoPRB2009}. 
Moreover, there is only a small barrier for the tetrahedral to migrate to an adjacent octahedral\cite{CarvalhoPRB2009}. 
The \intp{} assignment is unambiguously confirmed by the excellent agreement between the calculated EFG of $1.21\cdot10^{20} \ \mathrm{V} /\mathrm{m}^2$ in \Table{\ref{table:calculated_EFG}} with the experimental value at room temperature of $1.342(1)\cdot10^{20} \ \mathrm{V} /\mathrm{m}^2$.
The unrelaxed $2a$ site is symmetrically coordinated by six O$^{2-}$ ions. 
Axial lattice relaxation brings the interstitial cation closer to three anions, 
lowering the energy. This off-centre site is also consistent with emission channeling from other 
alkali radioisotopes in wurtzite crystals\cite{WahlSST2016,WahlPRL2017, WahlJAP2020}. 
We note the relaxed site is still 3-fold symmetric about the $c$ axis, so the EFG remains axial, 
consistent with Eq.\ (\ref{firstordersats}) for the satellite splittings. As a corollary, 
we then attribute the broad low temperature resonance in Fig.\ \ref{figure:300K_AG} that disappears by 300 K 
to \intp{} in the vicinity of some implantation-related crystalline disorder, 
also consistent with emission channeling in related materials\cite{WahlJAP2020}.

Next we consider the B site resonance which is well-resolved only in the comb spectra. 
Though it can be identified as low as 210 K, it only becomes substantial above 300 K,
increasing in amplitude towards the highest $T$.
This suggests B is the product of a site change transition to a lower energy site, 
likely the substitutional. This is similar to $^{24}$Na implanted into ZnO where emission channeling 
finds the interstitial twice as probable as the substitutional at room temperature, 
while by 420 K the interstitial is largely converted to substitutional\cite{WahlSST2016}. 
In contrast to the A site, the calculated EFG is substantially smaller than measured. 
However, Na \bnmr{} also finds two sites at room temperature, 
with the substitutional site having a factor of 2 smaller EFG than the interstitial\cite{MinamisonoNIMPR2010,WahlSST2016}, very similar to \eli. 
The evident underprediction of the EFG by DFT (see Table \ref{table:calculated_EFG}) is probably due to its sensitivity 
to the detailed lattice relaxation around the \subn{} which may not be captured accurately with the GGA functional.

The temperature dependence of the spectra reveals further aspects of the site change.
The A site line (most clearly revealed in the comb resonances in Fig.\ \ref{figure:comb_data}) increases in amplitude above 200 K, 
reaching a maximum near room temperature. Some of this increase is due to narrowing. 
The bi-Lorentzian lineshape also suggests two A sites sharing the same average EFG, 
but distinguished by the width of the EFG distribution. 
The broad component probably corresponds to the A site with some disorder at distances of a few lattice constants, 
while the narrow component probably represents an A site with ideal local structure. 
The growth of the narrow component is consistent with annealing of some correlated (small) Frenkel pairs in the vicinity. 
This agrees with optically detected electron paramagnetic resonance in electron irradiated 
ZnO that demonstrates low temperature annealing of the zinc (65-170 K) and oxide (160-230 K) sublattices \cite{VlasenkoPRB2005}. As seen clearly in Fig.\ \ref{figure:comb_full_spectrum}, 
above 300 K, the A line first broadens at 370 K, then narrows again by 400 K, 
maintaining nearly the same integrated area, while B increases substantially. 
If the site change involved an isolated interstitial becoming substitutional, 
e.g.\ by a knock out mechanism\cite{Knutsen2013}, then we would expect a decrease in the A amplitude with the simultaneous growth of B, 
such as in the well-established site change in the FCC metals\cite{MorrisPRL2004}. 
This is not the case, however. While it is possible that dynamic effects on the amplitude are confounding here, 
a careful inspection of the spectra in Fig.\ \ref{figure:comb_full_spectrum} reveals that the increase in 
B appears to be at the expense of the very broad background intensity that underlies the better resolved resonances. 
Production of \subn{} in this $T$ range then appears to result from a perturbed (disordered) site as a starting point rather than the isolated interstitial directly. 
This is reasonable for a site change involving a nearby vacancy (which is almost certainly implantation-related, 
since thermodynamic vacancies are extremely rare). Interstitials nearest to such vacancies will be the first to become substitutional, 
and they will also experience a significantly perturbed EFG in advance of the site change. 
In the ideal crystal, the barrier for Zn vacancy migration is quite high\cite{JanottiPRB2007}, 
so the site change is probably driven by mobility of the interstitial Li. 
We consider further the mobility of \intp{} below in relation to the spin-lattice relaxation.

Thus far, the \eli\ in ZnO follows the scenario typical for light implanted isotopes as outlined above. However, we also find clear evidence for another
site (denoted C) with a well-defined EFG in the
comb spectrum in Fig.\ \ref{figure:comb_full_spectrum}(a).
This site may correspond to a complex with another point defect, such as an oxygen vacancy,
Li$_{\mathrm{Zn}}-\mathrm{V}_\mathrm{O}$, which is Coulombically bound. This would
be consistent with the high temperature appearance of the closely
related Na$_{\mathrm{Zn}}-\mathrm{V}_\mathrm{O}$ complex seen by electron-nuclear double resonance \cite{OrlinskiiPRL2004}, 
but the fact that the EFG for C is even smaller than B seems inconsistent with a nearby charged vacancy.
Moreover, such a complex would have several orientations
relative to the applied field which should yield a more complex spectrum.
However, if the complex is {\it dynamic}, e.g.\ with $\mathrm{V}_\mathrm{O}$ hopping 
rapidly (on the scale of the quadrupolar splitting of several kHz) among the three equivalent neighbours and axial position of \subn{}, 
then a single reduced average EFG might result. This suggestion could be tested with more detailed calculations.

Finally, we consider the {\it widths} of the well-resolved resonances.
Without significant dipolar broadening, we might expect the resonances to be extremely narrow.
However, the widths in \Table{\ref{table:linewidth}} are considerably larger than
similarly split satellites in crystals of Bi\cite{wam2014} or {Sr$_2$RuO$_4$}\cite{Cortie2015}, 
probably due to local crystalline disorder\cite{JanottiPRB2007}. 
The primary broadening mechanism is quadrupolar, as confirmed by the substantially narrower MQ resonances.
Another potential source of broadening is the inhomogeneous magnetic fields due
to a population of dilute paramagnetic centres, which could be intrinsic defects\cite{Esquinazi2020} or impurities such as Fe. 
In NMR, this results in a term in the linewidth proportional to the impurity spin polarization that, 
in the dilute limit, is typically Curie-like (proportional to $1/T$). The absence of such a temperature dependent broadening is clear in Fig.\ \ref{figure:comb}(b), 
which shows the width to be quite independent of temperature below 150 K. 
The slight narrowing above this is probably related to the healing of nearby Frenkel pairs. 
This is consistent with predominantly quadrupolar broadening down to the lowest temperature. 
However, the MQ satellites, which are not quadrupole broadened, 
do exhibit a low $T$ broadening (see \Table{\ref{table:linewidth}}), 
suggesting a Curie term may be emerging at the lowest $T$. 
At the opposite end of the $T$ range, the broadening at 370 K is probably dynamic, 
and we discuss it below with the spin-lattice relaxation.

\subsection{Magnetic State of Implanted \eli}

In the previous section, we found very little evidence for broadening due to dilute 
paramagnetic defects. Here we consider paramagnetism in the immediate vicinity of the \eli.
It is widely accepted that Li is an amphoteric dopant in ZnO, where the \intLi\ is a donor
and \subLi\ an acceptor. The question then arises: what is the charge state?
In the ionized state, both would be diamagnetic, while un-ionized they would be paramagnetic
with an unpaired electron/hole spin localized in the vicinity.
This distinction should be dramatic in the data, since the
hyperfine interaction provides a strong magnetic perturbation on the NMR.
However, we see no evidence for a hyperfine splitting for either site.
This appears inconsistent with the EPR of \subLi\cite{Schirmer1968}.
However, the EPR signal requires cross gap photoexcitation of carriers, 
meaning it is actually due to a {\it metastable} paramagnetic defect. From our data, 
we conclude that \subLi\ is fully ionized in the temperature range where we observe it (above 200 K), meaning it is quite a shallow acceptor.
Similarly, \intLi\ remains diamagnetic over the entire range down to the lowest $T$ (7.6 K). 
Again this appears inconsistent with the EPR (and ENDOR) of the interstitial\cite{OrlinskiiPRL2004}, but 
this signal is observed (without photoexcitation) at 1.6~K, so it may be simply that ionization occurs in the intervening factor of 5 in $T$. 
However, this signal is also from nanoparticulate ZnO, and there is a strong size effect on the hyperfine coupling, 
due to confinement of the impurity wavefunction, which decreases strongly with increasing particle size. 
At 7.6 K, we place an upper bound of a hyperfine splitting at 2 kHz (smaller than 3 Gauss at the \eli\ nucleus), 
corresponding to a fraction of the MQ linewidths in \Table{\ref{table:linewidth}}, confirming that \intLi\ is shallow as a donor in the bulk.

While the carrier concentration in our samples is quite low and far below the metallic limit, 
here we consider the possibility that the Li defects are un-ionized but rapidly exchanging with a population of 
free carriers. In this case, the hyperfine splitting is averaged out, 
but there is a remnant time average shift of the NMR which typically follows the Curie law\cite{Meintjes2005,Chow2000}. 
The observed shift values in \Table{\ref{table:linewidth}} show no evidence for such a contribution, 
remaining small (in the range of Li chemical shifts) at all temperatures. This provides further confirmation that the observed Li is fully ionized.

Fluctuations of the hyperfine field of an unpaired electron in the vicinity 
may well cause the \eli\ nuclear spin to relax so rapidly that it is not observed. 
This might account for the small missing fraction in the AG crystal. 
If this is the case, its reduction in the annealed crystal suggest that
the probability of such a paramagnetic environment is reduced by annealing, 
i.e.\ it does not appear characteristic of pure ZnO. 

\subsection{Spin-Lattice Relaxation and Dynamics}

Spin-lattice relaxation is driven by fluctuations of the local fields at the nucleus, 
specifically, the Fourier component at the NMR frequency (41.27 MHz).
For quadrupolar nuclei, the most important contribution is usually EFG fluctuations, 
since this is the largest term in the nuclear spin Hamiltonian (after the Zeeman interaction with $B_0$).
This hierarchy of interaction strengths is reflected in the time average spectra above, where the
quadrupolar splitting far exceeds the satellite linewidths.

Below 300 K, the relaxation is so slow that it is difficult to measure reliably,
due to lifetime of the probe. In this temperature range, thermal phonons could cause SLR
by a Raman process, producing a $1/T_1$ varying approximately as $T^2$, e.g.\ in LaAlO$_3$,
see Ref.\ \onlinecite{KarnerPMSR2018}. Such phonon relaxation is evidently too weak to measure.

In contrast, with an onset around 300 K, there is an activated increase in the $1/T_1^{\mathrm{slow}}$ above room temperature. 
This could, in principle, be due to thermally excited carriers\cite{Selbach1979}, 
but the peak in $1/T_1^{\mathrm{slow}}$ in Fig.\ \ref{figure:SLR_rates} is not consistent with this. 
Such a peak is typical of some stochastic motion, either of the probe ion itself or some other species in its vicinity, 
giving rise to a fluctuating field at the nucleus.
In the simplest case of the isolated \intp{} hopping diffusively on the sublattice of octahedral sites, 
the EFG at each site is equivalent, both in magnitude and direction, 
so fluctuations would occur only as the ion briefly transits between sites. 
On the other hand, the fluctuating EFG may be due to another nearby interstitial ion, 
probably the more mobile Zn$_{\mathrm{I}}$, which is likely part of the nearest Frenkel pair. 
If the corresponding vacancy remains  nearby, then the dynamics of both interstitial cations 
would be modified by the Coulomb potential well of the negatively charged vacancy.
The $1/T_1^{\mathrm{slow}}$ maximum may then correspond to a Bloembergen Purcell Pound (BPP) peak\cite{BloembergenPR1948}, 
where a fluctuating field described
by a single exponential correlation time $\tau_c$ with an activated temperature 
dependence $\tau_0 \mathrm{exp}(E_a / k_bT)$ sweeps through the NMR frequency, 
producing a peak when $\tau_c$ matches $\omega_0^{-1}$. Fitting the measured $1/T_1$ to the BPP model produces the curve shown
in Fig.\ \ref{figure:SLR_rates} 
with fitted values of $E_a = 0.57(2)$ eV and $\tau_0 = 3(1) \cdot$ 10$^{-16}$ s.
While the fit is not too bad, there are several important features that are inconsistent with a 
simple BPP interpretation: 1) The fitted ``attempt frequency'' $1/\tau_0$ is several orders of magnitude higher than expected. 
This value is related to the narrowness of the peak in temperature.
2) The SLR peak coincides with a {\it broadening} in the resonances (see Fig.\ \ref{figure:comb}). 
In contrast, the spectrum should be motionally narrowed when $\tau_c^{-1}$ exceeds the linewidth 
(several orders of magnitude smaller than $\omega_0$), i.e.\ at temperatures well below the SLR peak.
3) Taking the $A$ site $\nu_q$ as the scale of fluctuations, the peak value of $1/T_1$ is also too low.
These inconsistencies suggest a more complicated situation than simple isolated interstitial diffusive motion.  

For this reason, we seek an alternate explanation. Consider the line broadening at high temperature. 
It is not magnetic as the MQ resonances are unaffected. Rather it resembles the broadening expected for an 
NMR probe fluctuating between different environments at a frequency comparable to the spectral difference of the two. 
The width is maximized simultaneously for both the A and B lines as well as the broad background in the comb spectra. 
From this we speculate that
some activated local dynamics starts to relax the \eli\ spin above 300 K, but before reaching the BPP peak, 
this relaxation channel is quenched. This might be \intp{} becoming mobile and then finding a vacancy and becoming the immobile \subn{}. 
On the other hand, it could also be Zn$_{\mathrm{I}}$ beginning to hop locally in the attractive potential of the vacancy, and then detrapping and moving away.
In either case, the activated dynamics is interrupted and the relaxation diminishes as this timescale becomes short enough to stop the relaxation.

\subsection{Temperature Dependence of the EFG}
\label{EFGT}

The electric field gradient at the nucleus is sensitively determined by the site symmetry and 
the local electronic structure about the implanted ion. It is not surprising then to find that it is temperature dependent. 
For the A site, it decreases by more than 10\% from its low $T$ value to 400 K, 
see Fig.\ \ref{figure:comb_data}. The corresponding values of $V_{zz}(T)$ shown in \Fig{\ref{figure:comb}}(c) were fit to
\begin{equation}
V_{zz}({T}) = V_{zz}(0)(1-{B}{T}^{\alpha}). 
\label{eq:EFG} 
\end{equation}
\noindent The resulting parameters are reported in \Table{\ref{table:efg}}, where a $T^{1.7}$ behavior is determined for an unconstrained fit, 
while fixing $\alpha=1.5$ produces a slightly worse fit with small deviations below 100 K.

\begin{table}[]
\centering
\caption[]{Fit Parameters from Eq.\ (\ref{eq:EFG}) for the temperature dependence of the EFG 
at the \lip\ A site in ZnO. $V_{zz}(0)$ is in units of $10^{20}\mathrm{V} /\mathrm{m}^2$. }
\label{table:efg}
\renewcommand*{\arraystretch}{1.4}
\begin{tabular}{ cccc } 
\hline
Fit type & $V_{zz}(0)$  & $B (\alpha)$  & $\alpha$  \\ 
\hline
\hline
Fixed $\alpha$  & 1.4507(2)   &  $1.47(2) \times 10^{-5}$  & 1.5 \\
Free $\alpha$   &  1.4428(3)    &  $4.4(2) \times 10^{-6}$   & 1.697(6)  \\ 
\hline
\end{tabular}
\end{table}

It is widely recognized that thermal expansion plays only a minor role in determining $V_{zz}(T)$
\cite{RaghavanPRB1976}. Instead, population of phonon modes
renders the EFG time dependent, and the time average value is reduced as the 
vibrational amplitude increases\cite{TorumbaPRB2006,NikolaevPRB2020}, 
resulting in an average behavior captured by the simple phenomenological form of Eq.\ (\ref{eq:EFG}), in a similar vein
to the temperature dependence of the energy gap in semiconductors\cite{Odonnell1991}.
The exponent $\alpha$ is often found to be 3/2 for implanted defects in non-cubic metals and 
metallic compounds\cite{ChristiansenZPBCM1976}, but, while common, it is by no 
means universal\cite{TorumbaPRB2006}. It's occurrence here demonstrates that it is not 
particular to metals\cite{NishiyamaPRL1976,ChristiansenZPBCM1976} or narrow gap semiconductors\cite{AmaralPLA1984}, 
nor is it characteristic of heavy isotope perturbed angular correlation probes.
It would be a strong confirmation of the {\it ab initio} models, 
if lattice dynamic calculations\cite{NikolaevPRB2020,TorumbaPRB2006} could reproduce the observed $V_{zz}(T)$.

In contrast, site B exhibits an anomalously increasing EFG, which, while rare, is not unheard of\cite{NikolaevPRB2020}. 
However, this should be interpreted with some caution, because
the dynamic broadening at 370 K may also affect the apparent $\nu_q$ of the resonances.
Confirmation of this behavior must await more detailed future measurements.

\section{Summary}

Using ion-implanted \eli\ \bnmr\, we characterized the microscopic dynamics, stability, and magnetic states of Li defects in high purity ZnO single crystals. 
RF comb measurements, through a strong signal enhancement, enabled identification of three distinct defects between 7.6 and 400 K. 
Comparison to DFT calculations confirm the stability of ionized shallow donor \intp\ up to 400 K and 
coexistence with \subn\ confirming the amphoteric character responsible for self-compensation of $p$-type doping. 
From the absence of a resonance shift or hyperfine splitting, we find no evidence for localized holes near \subLi\ down to 210 K, 
indicating it is a shallow acceptor in isolated form. Above 370 K, a third unexpected defect is detected that we tentatively suggest to be a complex with \subn. 
From spin-lattice relaxation measurements, a $T$ independent fraction of the initial polarization suggest 
some \eli\ stop near magnetic defects with unpaired electrons that are reduced in concentration by annealing. 
The spin-lattice relaxation rate peaks at 370 K. We ascribe its temperature dependence to the coupled motional dynamics of Li and other nearby defects. 

Our results using ultradilute \eli{}\ reveal the defect properties of Li in ZnO and indicate while $p$-type doping is possible with Li, 
self-compensation by interstitial Li poses a barrier. Characterization of the local \eli\ environment is a first step toward studies on surface properties, 
intentionally doped crystals, and the structurally related but more complex thin-film transistor material, IGZO.
Finally, our measurements of $V_{zz}(T)$ for \intLi\ demonstrating a $T^{1.7}$ dependence resembling a 
non-cubic metal provides a case to validate the \textit{ab initio} lattice dynamics 
models for a light interstitial in a semiconductor \cite{NikolaevPRB2020,TorumbaPRB2006}.

\begin{acknowledgments}
We thank K. Foyevtsova and I. Elfimov for assistance with the DFT calculations; 
S. Daviel and H. Hui of TRIUMF for implementation of the RF comb;
M. McLay and S. Chan for the high temperature detector upgrade; 
Y. Cai for help with measurements;
Useful discussions: J. St{\"a}hler. 
This work was supported by NSERC Discovery.
D.F., V.L.K., and J.O.T. acknowledge the additional support from their SBQMI QuEST fellowship. 
\end{acknowledgments}
 
\appendix
\section{The Fast Relaxing Component}
\label{fastcomp}

\begin{figure}[ht]
\begin{center}
\includegraphics[width=\columnwidth]{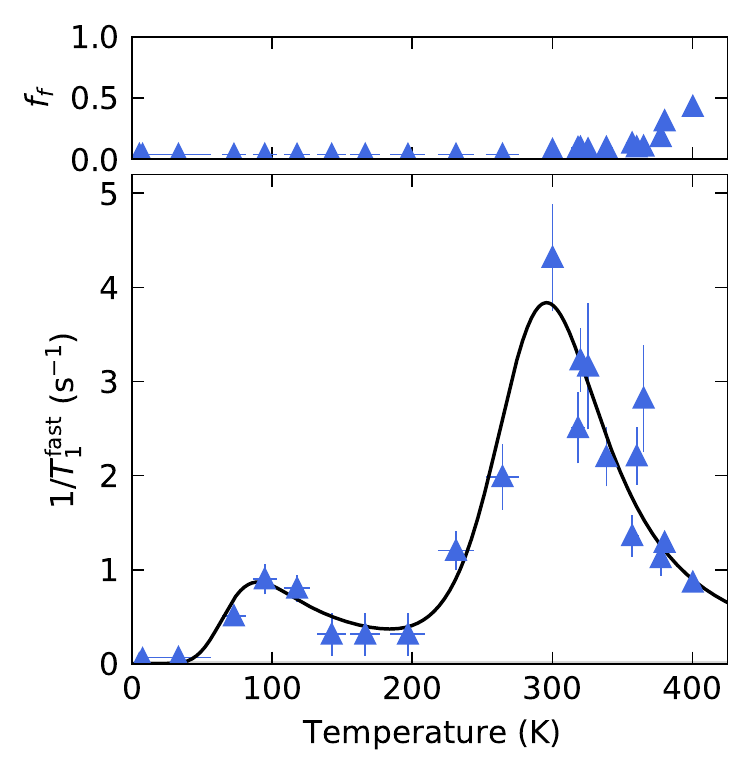}
\caption{The fast component's spin-lattice relaxation rate, $1/T_1^{\mathrm{fast}}$ from biexponential fits to 
the data as shown in Fig.\ \ref{figure:SLR_rates_fast} at 6.55 T $\parallel$ c. 
The peak occurs at about 300 K, significantly lower than the slow component.}
\label{figure:SLR_rates_fast} 
\end{center}
\end{figure}

As discussed above, we fit the spin-lattice relaxation data to a biexponential relaxation function.
Over much of the temperature range, the fast component ($f_f = 1-f_s$) is small, 
accounting for only a few percent of the signal. As a result, it is much less well-determined than the larger slow relaxing component.
For this reason, we did not discuss it in the main text. We include it here for completeness.
\Fig{\ref{figure:SLR_rates_fast}} shows its rate and fraction.

For a spin-2 nucleus relaxing by slow quadrupolar fluctuations, 
one expects a biexponential with a fast component of about this amplitude\cite{BeckerJPS1982}. 
However, the observed signal deviates from this model in several significant ways: 1) $\lambda_f$ is substantially 
faster relative to $\lambda_s$ than predicted; 2) $\lambda_f(T)$ should track $\lambda_s(T)$, 
while \Fig{\ref{figure:SLR_rates_fast}} shows that the main peak is substantially lower in temperature, 
and there is a secondary peak around 100 K; 3) When the fluctuations become fast, 
above the $T_1$ minimum, the fast amplitude should go smoothly to zero. Instead, it grows in this region.
For these reasons, the fast component probably has a different origin. 
Below 350 K, where its amplitude is small, it may correspond to \eli\ in some exceptional environment 
with a significantly higher local relaxation rate, perhaps near the surface, 
but more measurements are required to make any firm conclusions.

\section{Stopping Distribution} 
\label{appendix_stopping}

\begin{figure}[ht]
\begin{center}
\includegraphics[width=\columnwidth]{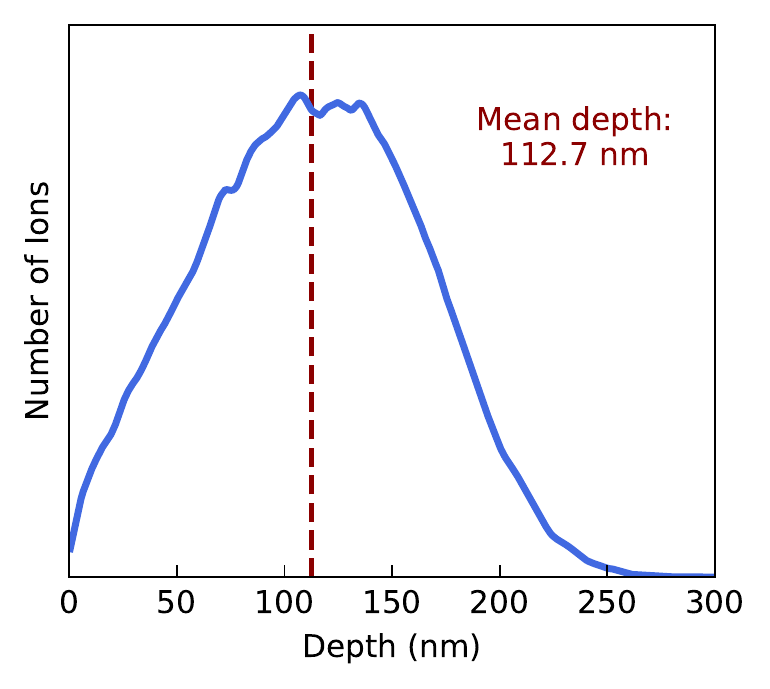}
\caption{SRIM\cite{ZieglerNIMP2010} stopping profile for 25 keV \elip\ implanted in ZnO.}
\label{figure:SRIM} 
\end{center}
\end{figure}

The energy of the \elip{} beam determines its implantation profile. 
We simulated this for 25 keV and 10$^5$ ions normally incident on ZnO with the SRIM Monte Carlo code, 
calculating a mean depth of 112.7 nm and a straggle (standard deviation) of 50.8 nm. 
The stopping distribution is shown in \Fig{\ref{figure:SRIM}}, indicating that the vast majority of \elip\ are well beyond the surface region 
where the electronic properties are modified ($\sim 2$ nm)\cite{Deinert2015}. 
The simulation does not take into account implantation channeling which would result 
in a tail to the profile towards larger depths due to the channeled fraction. 

\section{Helicity-Resolved Resonance Spectra} 
\label{appendix_resonance}

\begin{figure}[hb]
\begin{center}
\includegraphics[width=\columnwidth]{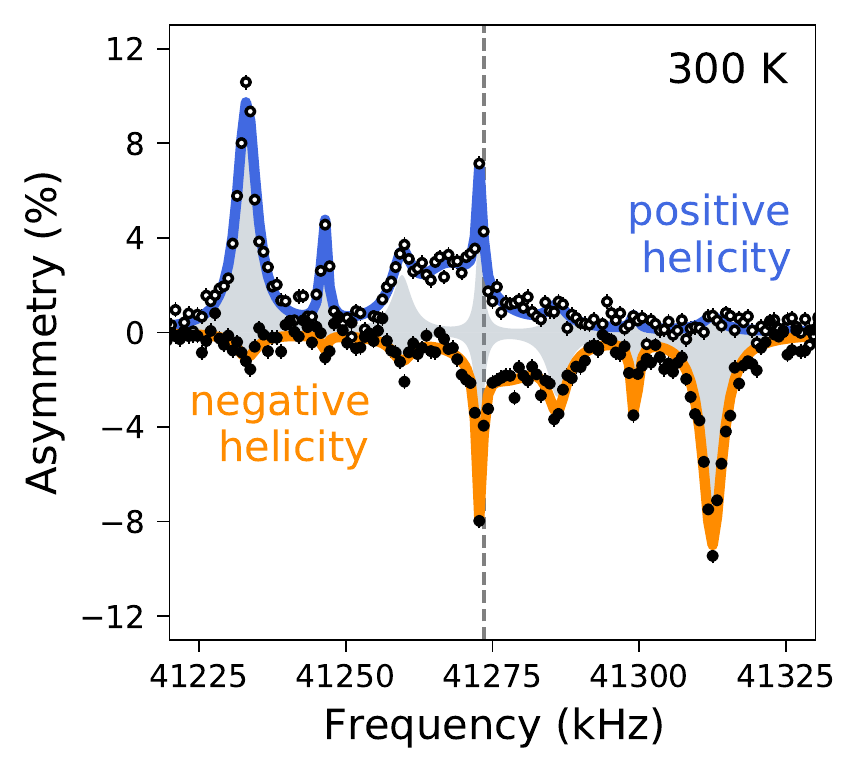}
\caption{The single RF spectra for \eli{} in the as-grown ZnO with 6.55 T$\parallel c$ at 300 K 
showing the two helicities: initial spin state primarily $m=2$ (open) and $m=-2$ (closed). 
The asymmetry is normalized to its off-resonance equilibrium value. The site A spectrum is shaded grey.}
\label{figure:helicity_resolved} 
\end{center}
\end{figure}

In Figure \ref{figure:helicity_resolved}, we present the single tone RF spectrum at 300 K, 
separately for the two helicities. The occurrence of corresponding satellites on opposite 
sides of the centre of the pattern in the two helicities is unambiguous confirmation that the splittings are quadrupolar.
The corresponding helicity combined spectrum is shown in Fig.\ \ref{figure:300K_AG}.

\section{Spin-Lattice Relaxation of the Annealed Crystal} 
\label{appendix_SLR}

\begin{figure}[ht]
\begin{center}
\includegraphics[width=\columnwidth]{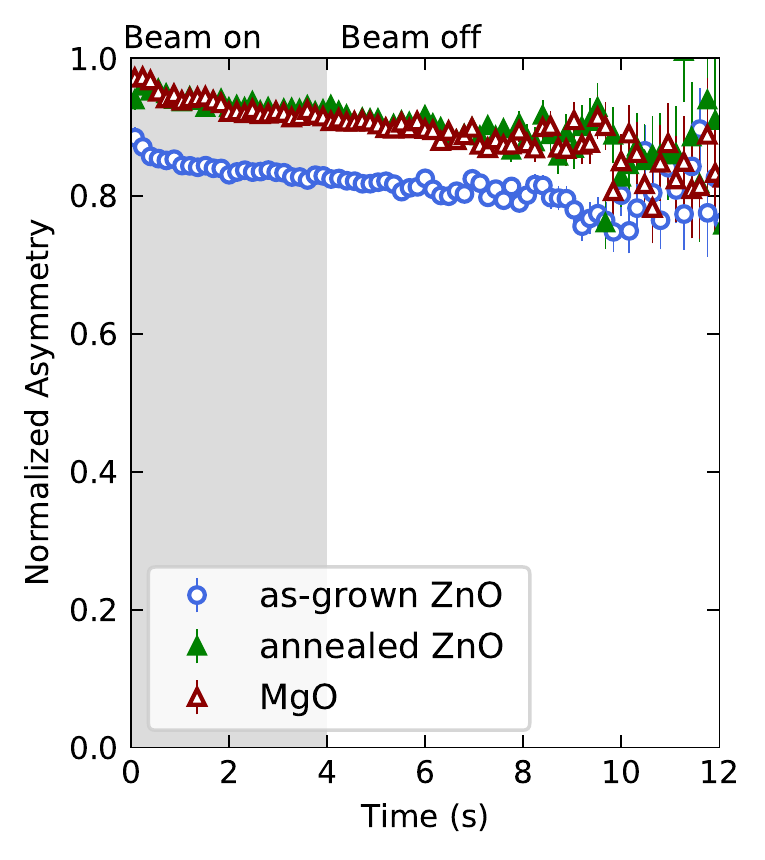}
\caption{The time dependence of the asymmetry for \elip\ in the as grown ZnO, annealed ZnO, and MgO reference at 300 K and 6.55 Tesla$\parallel c$. 
The decay is due to spin-lattice relaxation of the isolated implanted \elip. 
The asymmetry is normalized to the initial asymmetry $A_0$ in MgO extracted from fitting a biexponential relaxation function. 
The as grown ZnO has a large reduction in the initial asymmetry, corresponding to a missing fraction.}
\label{figure:SLR_annealed} 
\end{center}
\end{figure}

In \Fig{\ref{figure:SLR_annealed}} we present a comparison of the NMR recovery curves at 300 K in annealed ZnO, 
as-grown ZnO, and the MgO reference. Using Eq.\ (\ref{eq:relaxation_fcn}) \cite{MacFarlaneJPCS2014} to fit the data, 
the initial asymmetry $A_0$ is extracted for \np\ ZnO and MgO. For MgO $A_0 = 0.095(1)$, 
while \np\ ZnO measured with the standard forward $\beta$ detector yielded $A_0 = 0.0864(8)$. 

\bibliographystyle{apsrev4-1}
\bibliography{references.bib}

\end{document}